\documentclass[envcountsect]{llncs}
\usepackage{multirow}
\usepackage[table,xcdraw]{xcolor}
\usepackage{enumitem}
\usepackage[pdftex]{graphicx}
\graphicspath{{figures/}}
\DeclareGraphicsExtensions{.jpg,.png}

\usepackage[caption=false,font=footnotesize]{subfig}

\usepackage{amsmath}
\usepackage{amssymb}
\usepackage{bbm}
\usepackage{stmaryrd}

\usepackage{array}
\usepackage{color}
\usepackage[hyphens]{url}
\usepackage[pdftitle=Title,pdfauthor=Anonymous]{hyperref}

\usepackage{xspace}
\newcommand{\etal}{\textit{et al.}\xspace}
\newcommand{\etc}{\textit{etc.}\xspace}
\newcommand{\ie}{\textit{i.e.,}\xspace}
\newcommand{\eg}{\textit{e.g.,}\xspace}
\newcommand{\cf}{\textit{cf.}\xspace}

\usepackage{adjustbox}
\newcommand{\headrow}[1]{\multicolumn{1}{c}{\adjustbox{angle=45,lap=\width-0.5em}{#1}}}

\newcommand{\sys}{\textsf{TokenHook}\xspace}
\newcommand{\erc}{ERC-20\xspace}
\newcommand{\num}{82\xspace}
\newcommand{\prct}{99.5\%\xspace}

\begin{document}
\frontmatter
\mainmatter

\title{\Large \bf \sys : Secure \erc smart contract}
\author{Reza Rahimian, Jeremy Clark}
\institute{Concordia University}

\maketitle



\begin{abstract}
\erc is the most prominent Ethereum standard for fungible tokens. Tokens implementing the \erc interface can interoperate with a large number of already deployed internet-based services and Ethereum-based smart contracts. In recent years, security vulnerabilities in \erc have received special attention due to their widespread use and increased value. We systemize these vulnerabilities and their applicability to \erc tokens, which has not been done before. Next, we use our domain expertise to provide a new implementation of the \erc interface that is freely available in Vyper and Solidity, and has enhanced security properties and stronger compliance with best practices compared to the sole surviving reference implementation (from OpenZeppelin) in the \erc specification. Finally, we use our implementation to study the effectiveness of seven static analysis tools, designed for general smart contracts, for identifying ERC-20 specific vulnerabilities. We find large inconsistencies across the tools and a high number of false positives which shows there is room for further improvement of these tools. 
\end{abstract}



\section{Introduction}\label{sect:introduction}
The Ethereum blockchain~\cite{EthGit,EIP150} allows users to build and deploy decentralized applications (DApps) that can accept and use its protocol-level cryptocurrency ETH. Many DApps also issue or use custom tokens. Such tokens could be financial products, in-house currencies, voting rights for DApp governance, or other valuable assets. To encourage interoperability with other DApps and web applications (exchanges, wallets, \etc), the Ethereum community accepted a popular token standard (specifically for fungible tokens) called \erc~\cite{ERC20Std}. While numerous \erc extensions or replacements have been proposed, \erc remains prominent. Of the 2.5M~\cite{Alethio} smart contracts on the Ethereum network, 260K are tokens~\cite{TokenTracker} and 98\% of these tokens are \erc~\cite{EtherScan}. 

The development of smart contracts has been proven to be error-prone, and as a result, smart contracts are often riddled with security vulnerabilities. An early study in 2016 found that 45\% of smart contracts at that time had vulnerabilities~\cite{MakSm}. \erc tokens are subset of smart contracts and security is particularly important given that many tokens have considerable market capitalization (\eg USDT, BNB, UNI, DAI, \etc). As tokens can be held by commercial firms, in addition to individuals, and firms need audited financial statements in certain circumstances, the correctness of the contract issuing the tokens is now in the purview of professional auditors. Later, we examine one static anaylsis tool from a `big-four' auditing firm.

\subsubsection*{Contributions} Ethereum has undergone numerous security attacks that have collectively caused more than US\$100M in financial losses~\cite{DAO1,PeckShield,PartiyMultiSig,MyEthWallet,ParityFirstHack,ParitySecondHack}. Although research has been done on smart contract vulnerabilities in the past~\cite{EthSecServ}, we focus specifically on \erc tokens. 

\begin{enumerate}
\item We study all known vulnerabilities and cross-check their relevance to \erc token contracts, systematizing a comprehensive set of \num distinct vulnerabilities and best practices. 
\item While not strictly a research contribution, we believe that our newly acquired specialized domain knowledge should be put to use. Thus, we provide a new \erc implementation, \sys, that is open source and freely available in both Vyper and Solidity.
\item \sys is positioned to increase software diversity: currently, no Vyper \erc implementation is considered a reference implementation, and only one Solidity implementation is actively maintained (OpenZeppelin's~\cite{OpenZepplin}). Relative to this implementation, \sys has enhanced security properties and stronger compliance with best practices. 
\item Perhaps of independent interest, we report on differences between Vyper and Solidity when implementing the same contract. \item We use \sys as a benchmark implementation to explore the completeness and precision of seven auditing tools that are widely used in industry to detect security vulnerabilities. We conclude that while these tools are better than nothing, they do not replace the role of a security expert in developing and reviewing smart contract code.
\end{enumerate}


\section{Sample of high profile vulnerabilities}\label{sec:vul}
In this section, we examine general attack vectors and cross-check their applicability to \erc tokens. We sample some high profile vulnerabilities, typically ones that have been exploited in real world \erc tokens\cite{SolidtySecBlog,EthSecServ,SoliditySecCon,ConsensysSecCon,LandoKL}. For each, we (i) briefly explain technical details, (ii) the ability to affect \erc tokens, and (iii) discuss mitigation techniques. Later we will compile a more comprehensive list of \num vulnerabilities and best practices (see Table\ref{tab:result1}), including these, however space will not permit us to discuss each one at the same level of detail as the ones we highlight in this section (however we will include a simple statement describing the issue and the mitigation).

\subsection{Multiple withdrawal}\label{subsec:mwa}
This \erc-specific issue was originally raised in 2017~\cite{MikVlad,TomHale}. It can be considered as a \textit{transaction-ordering}~\cite{OrderingAttack} or \textit{front-running}~\cite{eskandari2019sok} attack. There are two \erc functions (\ie \texttt{Approve()} and \texttt{transferFrom()}) that can be used to authorize a third party for transferring tokens on behalf of someone else. Using these functions in an undesirable situation (\ie front-running or race-condition) can result in allowing a malicious authorized entity to transfer more tokens than the owner wanted. There are several suggestions to extend \erc standard (\eg MonolithDAO~\cite{MonolithDAO} and its extension in OpenZeppelin~\cite{OpenZepplin}) by adding new functions (\ie \texttt{decreaseApproval()} and \texttt{increaseApproval()}), however, securing \texttt{transferFrom()} method is the effective one while adhering specifications of the \erc standard~\cite{ERC20MWA}.

\subsection{Arithmetic Over/Under Flows.}\label{subsec:ovf}
An \textit{integer overflow} is a well known issue in many programming languages. For \erc, one notable exploit was in April 2018 that targeted the BEC Token~\cite{BECToken} and resulted in some exchanges (\eg OKEx, Poloniex, \etc) suspending deposits and withdrawals of all tokens. Although BEC developers had considered most of the security measurements, only line 261 was vulnerable~\cite{Osiris,PeckShield}. The attacker was able to pass a combination of input values to transfer large amount of tokens~\cite{Overflow}. It was even larger than the initial supply of the token, allowing the attacker to take control of token financing and manipulate the price. In Solidity, integer overflows do not throw an exception at runtime. This is by design and can be prevented by using the \texttt{SafeMath} library~\cite{SafeMath} wherein \texttt{a+b} will be replaced by \texttt{a.add(b)} and throws an exception in the case of arithmetic overflow. Vyper has built-in support for this issue and no need to use \texttt{SafeMath} library.

\subsection{Re-entrancy}\label{subsec:ent}
One of the most studied vulnerabilities is re-entrancy, which resulted in a US\$50M attack on a DApp (called the DAO) in 2016 and triggered an Ethereum hard-fork to revert~\cite{DAO1}. At first glance, re-entrancy might seem inapplicable to \erc however any function that changes internal state, such as balances, need to be checked. Further, some \erc extensions could also be problematic. One example is ORBT tokens~\cite{ORBTToken} which support token exchange with ETH without going through a crypto-exchange~\cite{ORBT}: an attacker can call the exchange function to sell the token and get back equivalent in ETH. However, if the ETH is transferred in a vulnerable way before reaching the end of the function and updating the balances, control is transferred to the attacker receiving the funds and the same function could be invoked over and over again within the limits of a single transaction, draining excessive ETH from the token contract. This variant of the attack is known as \textit{same-function re-entrancy}, but it has three other variants: \textit{cross-function}, \textit{delegated} and \textit{create-based} ~\cite{SEREUM}. Mutex~\cite{WiKiMutex} and CEI~\cite{SolidtyDocSec} techniques can be used to prevent it. In Mutex, a state variable is used to lock/unlock transferred ETH by the lock owner (\ie token contract). The lock variable fails subsequent calls until finishing the first call and changing requester balance. CEI updates the requester balance before transferring any fund. All interactions (\ie external calls) happen at the end of the function and prevents recursive calls. Although CEI does not require a state variable and consumes less Gas, developers must be careful enough to update balances before external calls. Mutex is more efficient and blocks \textit{cross-function} attack at the beginning of the function regardless of internal update sequences. CEI can also be considered as a best practice and basic mitigation for the \textit{same-function re-entrancy}. We implement a \texttt{sell()} and \texttt{buy()} function in \sys for exchanging between tokens and ETH. \texttt{sell()} allows token holders to exchange tokens for ETH and \texttt{buy()} accepts ETH by adjusting buyer's token balance. It is used to buy and sell tokens at a fixed price (\eg an initial coin offering (ICO), prediction market portfolios~\cite{CBN+14}) independent of crypto-exchanges, which introduce a delay (for the token to be listed) and fees. Both CEI and Mutex are used in \sys to mitigate two variants of re-entrancy attack.

\subsection{Unchecked return values}\label{subsec:urv}
In Solidity, sending ETH to external addresses is supported by three options: \texttt{call.value()}, \texttt{transfer()}, or \texttt{send()}. The \texttt{transfer()} method reverts all changes if the external call fails, while the other two return a boolean value and manual check is required to revert transaction to the initial state~\cite{SoliditySendEther}. Before the \textit{Istanbul} hard-fork~\cite{IstanbulUpgrades}, \texttt{transfer()} was the preferred way of sending ETH. It mitigates reentry by ensuring ETH recipients would not have enough gas (\ie a 2300 limit) to do anything meaningful beyond logging the transfer when execution control was passed to them. EIP-1884~\cite{EIP1884} has increased the gas cost of some opcodes that causes issues with \texttt{transfer()}\footnote{After \textit{Istanbul}, the \texttt{fallback()} function consumes more than 2300 Gas if called via \texttt{transfer()} or \texttt{send()} methods.}. This has led to community advice to use \texttt{call.value()} and rely on one of the above re-entrancy mitigations (\ie Mutex or CEI)~\cite{WiKiMutex,CEI}. This issue is addresses in Vyper and there is no need to check return value of \texttt{send()} function.

\subsection{Frozen Ether}\label{subsec:feth}
As \erc tokens can receive and hold ETH, just like a user accounts, functions need to be defined to withdraw deposited ETH (including unexpected ETH). If these functions are not defined correctly, an \erc token might hold ETH with no way of recovering it (\cf Parity Wallet~\cite{ParityWalletHack}). If necessary, developers can require multiple signatures to withdraw ETH.

\subsection{Unprotected Ether Withdrawal}\label{subsec:uew}
Improper access control may allow unauthorized persons to withdraw ETH from smart contracts (\cf Rubixi~\cite{Rubixi}). Therefore, withdrawals must be triggered by only authorized accounts and ideally multiple parties.

\subsection{State variable manipulation}\label{subsec:svm}
The \texttt{DELEGATECALL} opcode enables a DApp to invoke external functions of other DApps and execute them in the context of calling contract (\ie the invoked function can modify the state variables of the caller). This makes it possible to deploy libraries once and reuse the code in different contracts. However, the ability to manipulate internal state variables by external functions has lead to incidents where the entire contract was hijacked (\cf the second hack of Parity MultiSig Wallet~\cite{ParitySecondHack}). Preventive techniques is to use \texttt{Library} keyword in Solidity to force the code to be stateless, where data is passed as inputs to functions and passed back as outputs and no internal storage is permitted~\cite{LIB1}. There are two types of Library: \textit{Embedded} and \textit{Linked}. Embedded libraries have only internal functions (EVM uses \texttt{JUMP} opcode instead of \texttt{DELEGATECALL}), in contrast to linked libraries that have public or external functions (EVM initiate a ``message call''). Deployment of linked libraries generates a unique address on the blockchain while the code of embedded libraries will be added to the contract's code ~\cite{LIB2}. It is recommended to use Embedded libraries to mitigate this attack.

\subsection{Balance manipulation}\label{subsec:blman}
\erc tokens generally receive ETH via a \textit{payable} function~\cite{SolidityDocPayable} (\ie \texttt{receive()}, \texttt{fallback()}, \etc), however, it is possible to send ETH without triggering payable functions, for example via \texttt{selfdestruct()} that is initiated by another contract~\cite{SolidityByExampleSelfDestruct}. This can cause an oversight where \erc may not properly account for the amount of ETH they have received~\cite{UnexpectedEth}. For example, A contract might use ETH balance to calculate exchange rate dynamically. Forcing ETH by attacker may affect calculations and get lower exchange rate. To fortify this vulnerability, contract logic should avoid using exact values of the contract balance and keep track of the known deposited ETH by a new state variable. Although we use \texttt{address(this).balance} in \sys, we do not check the exact value of it (\ie \texttt{address(this).balance == 0.5 ether)}---we only check whether the contract has enough ETH to send out or not. Therefore, there is no need to use a new state variable and consume more Gas to track contract's ETH. However, for developers who need to track it manually, we provide \texttt{contractBalance} variable. Two complementary functions are also considered to get current contract balance and check unexpected received ETH (\ie \texttt{getContractBalance()} and \texttt{unexpectedEther()}).

\subsection{Public visibility}\label{subsec:pvis}
In Solidity, visibility of functions are \texttt{Public} by default and they can be called by any external user/contract. In the Parity MultiSig Wallet hack~\cite{ParityFirstHack}, an attacker was able to call public functions and reset the ownership address of the contract, triggering a \$31M USD theft. It is recommended to explicitly specify visibility of functions instead of default \texttt{Public} visibility.


\section{A sample of best practices}\label{sec:bp}

We highlight a few best practices for developing DApps. Some best practices are specific to \erc, while others are generic for all DApps---in which case, we discuss their relevance to \erc.

\subsection{Compliance with \erc.}\label{subsec:compl}
According to the \erc specifications, all six methods and two events must be implemented and are not optional. Tokens that do not implement all methods (\eg GNT which does not implement the \texttt{approve()}, \texttt{allowance()} and \texttt{transferFrom()} functions due to \textit{front-running}\cite{GNT}) can cause failed function calls from other applications. They might also be vulnerable to complex attacks (\eg Fake deposit vulnerability\cite{DEPOSafe}, Missing return value bug\cite{ErcBug}).

\subsection{External visibility.}\label{subsec:external}
Solidity supports two types of \textit{function calls}: internal and external~\cite{SolidityDoc}. Note that functions calls are different than functions visibility (\ie Public, Private, Internal and External) which confusingly uses overlapping terminology. Internal function calls expect arguments to be in memory and the EVM copies the arguments to memory. Internal calls use \texttt{JUMP} opcodes instead of creating an \textit{EVM call}.\footnote{Also known as ``message call'' when a contract calls a function of another contract.} Conversely, External function calls create an \textit{EVM call} and can read arguments directly from the \texttt{calldata} space. This is cheaper than allocating new memory and designed as a read-only byte-addressable space where the data parameter of a transaction or call is held\cite{EthInDepth}. A best practice is to use external visibility when we expect that functions will be called externally.

\subsection{Fail-Safe Mode.}\label{subsec:failsf}
In the case of a detected anomaly or attack on a deployed \erc token, the functionality of the token can be frozen pending further investigation. For regulated tokens, the ability for a regulator to issue a `cease trade' order is also generally required. 

\subsection{Firing events.}\label{subsec:evnts}
In \erc standard, there are two defined events: \texttt{Approval} and \texttt{Transfer}. The first event logs successful allowance changes by token holders and the second logs successful token transfers by the \texttt{transfer()} and \texttt{transferFrom()}. These two events must be fired to notify external application on occurred changes. The external application (\eg TokenScope\cite{TokenScope}) might use them to detect inconsistent behaviors, update balances, show UI notifications, or to check new token approvals. It is a best practice to fire an event for every state variable change.

\subsection{Global or Miner controlled variables.}\label{subsec:miner}
Since malicious miners have the ability to manipulate global Solidity variables (\eg \texttt{block.timestamp}, \texttt{block.number}, \texttt{block.difficulty}, \etc), it is recommended to avoid these variables in \erc tokens.

\subsection{Proxy contracts.}\label{subsec:prxy}
An \erc token can be deployed with a pair of contracts: a proxy contract that passes through all the function calls to a second functioning \erc contract\cite{ProxyContract,ProxyPatterns}. One use of proxy contract is when upgrades are required---a new functional contract can be deployed and the proxy is modified to point at the update. Form audit point of view, it is recommended to have non-upgradable \erc tokens. 

\subsection{DoS with Unexpected revert.}\label{subsec:rvt}
A function that attempts to complete many operations that individually may revert could deadlock if one operation always fails. For example, \texttt{transfer()} can throw an exception---if one transfer in a sequence fails, the whole sequence fails. One standard practice is to account for ETH owed and require withdrawals through a dedicated function. In \sys, ETH is only transferred to a single party in a single function \texttt{sell()}. It seems overkill to implement a whole accounting system for this. As a consequence, a seller that is incapable of receiving ETH (\eg operating from a contract that is not payable) will be unable to sell their tokens for ETH. However they can recover by transferring the tokens to a new address to sell from. 

\subsection{Unprotected SELFDESTRUCT}\label{subsec:slfd}
Another vulnerability stemming from the second Parity wallet attack~\cite{ParitySecondHack} is protecting the \texttt{SELFDESTRUCT} opcode which removes a contract from Ethereum. The self-destruct method is used to kill the contract and its associated storage. \erc tokens should not contain \texttt{SELFDESTRUCT} opcode unless there is a multi approval mechanism.

\subsection{DoS with block gas limit.}\label{subsec:glimit}
The use of loops in contracts is not efficient and requires considerable amount of Gas to execute. It might also cause DoS attack since blocks has a \textit{Gas limit}. If execution of a function exceeds the block gas limit, all transactions in that block will fail. Hence, it is recommended to not use loops and rely on \texttt{mappings} variables in \erc tokens. 


\begin{figure}[t!]
	\centering
	\includegraphics[width=1.0\linewidth]{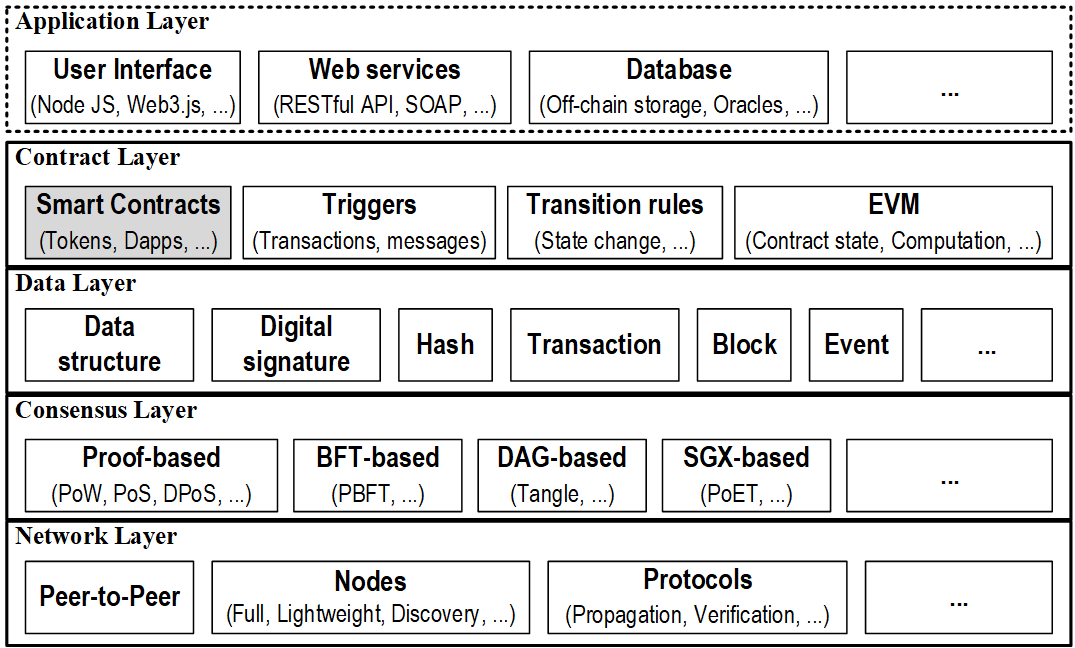}
	\caption{Architecture of the Ethereum blockchain in layers, including the interactive environment (\ie application layer). \erc tokens falls under the \textit{Smart Contracts} category in \textit{Contract Layer}.}\label{fig:blockchain}
\end{figure}

\section{\sys}\label{sec:proposal}
\sys is our ERC20-compliant implementation written in Vyper (v. 0.2.8) and Solidity (v. 0.8.4) \footnote{\sys deployed on Rinkeby at \url{https://bit.ly/33wDENx} (Solidity) and \url{https://bit.ly/3dXaaPc} (Vyper). Mainnet at \url{https://bit.ly/35FMbAf} (Solidity 0.5.11)}. It can be customized by developers, who can refer to each mitigation technique separately and address specific attacks. The presence of security vulnerability in supplementary layers (\ie consensus, data, network. \etc) affect the entire Ethereum blockchain, not necessarily \erc tokens. Therefore, vulnerabilities in other layers are assumed to be out of the scope. Required comments have been also added to clarify the usage of each function. Standard functionalities of the token (\ie \texttt{approve()}, \texttt{transfer()}, \texttt{transferFrom()}, \etc) have been unit tested. A demonstration of token interactions and event triggering can also be seen on Etherscan.\footnote{Etherscan: \url{https://bit.ly/33xHfL2}, \url{https://bit.ly/35TimMW} and \url{https://bit.ly/3eFAnAZ}} 

Among the layers of the Ethereum blockchain, \erc tokens fall under the \textit{Contract layer} in which DApps are executed. The presence of a security vulnerability in supplementary layers affect the entire Ethereum blockchain, not necessarily \erc tokens. Therefore, vulnerabilities in other layers are assumed to be out of the scope. (\eg \textit{Indistinguishable chains} at the data layer, the \textit{51\% attack} at the consensus layer, \textit{Unlimited nodes creation} at network layer, and \textit{Web3.js Arbitrary File Write} at application layer). 

Moreover, we exclude vulnerabilities identified in now outdated compiler versions. Examples: \textit{Constructor name ambiguity} in versions before 0.4.22, \textit{Uninitialized storage pointer} in versions before 0.5.0, \textit{Function default visibility} in versions before 0.5.0, \textit{Typographical error} in versions before 0.5.8, \textit{Deprecated solidity functions} in versions before 0.4.25, \textit{Assert Violation} in versions before 0.4.10, \textit{Under-priced DoS attack} before EIP-150 \& EIP-1884).

\subsection{Security features}
In our research, we developed \num security vulnerabilities and best practices for \erc. We concentrate here on how \sys mitigates these attacks. While many of these attacks are no doubt very familiar to the reader, our emphasis is on their relevance to \erc.

\subsubsection{Multiple Withdrawal Attack}
Without our counter-measure, an attacker can use a front-running attack~\cite{OrderingAttack,eskandari2019sok} to transfer more tokens than what is intended (approved) by the token holder. We secure the \texttt{transferFrom()} function by tracking transferred tokens to mitigate the \textit{multiple withdrawal} attack~\cite{ERC20MWA}. Securing the \texttt{transferFrom()} function is fully compliant with the \erc standard without the need of introducing new functions such as \texttt{decreaseApproval()} and \texttt{increaseApproval()}. 

\subsubsection{Arithmetic Over/Under Flows} 
In Solidity implementation, we use the \texttt{SafeMath} library in all arithmetic operations to catch over/under flows. Using it in Vyper is not required due to built-in checks.

\subsubsection{Re-entrancy} 
At first glance, re-entrancy might seem inapplicable to \erc. However any function that changes internal state, such as balances, need to be checked. We use Checks-Effects-Interactions pattern (CEI)~\cite{CEI} in both Vyper and Solidity implementations to mitigate \textit{same-function re-entrancy} attack. Mutual exclusion (Mutex)~\cite{WiKiMutex} is also used to address \textit{cross-function re-entrancy} attack. Vyper supports Mutex by adding \texttt{@nonreentrant(<key>)} decorator on a function and we use \texttt{noReentrancy} modifier in Solidity to apply Mutex. Therefore, both re-entrancy variants are addressed in \sys. 
 
\subsubsection{Unchecked return values}
Unlike built-in support in Vyper, we must check the return value of \texttt{call.value()} in Solidity to revert failed fund transfers. It mitigates the \textit{unchecked return values} attack while making the token contract compatible with EIP-1884~\cite{EIP1884}. 

\subsubsection{Frozen Ether}
We mitigate this issue by defining a \texttt{withdraw()} function that allows the owner to transfer all Ether out of the token contract. Otherwise, unexpected Ether forced onto the token contract (\eg from another contract running \texttt{selfdestruct}) will be stuck forever. 

\subsubsection{Unprotected Ether Withdrawal}
We enforce authentication before transferring any funds out of the contract to mitigate \textit{unprotected Ether withdrawal}. Explicit check is added to the Vyper code and \texttt{onlyOwner} modifier is used in Solidity implementation. It allows only owner to call \texttt{withdraw()} function and protects unauthorized Ether withdrawals. 

\subsubsection{State variable manipulation}
In the Solidity implementation, we use embedded \texttt{Library} code (for \texttt{SafeMath}) to avoid external calls and mitigate the \textit{state variable manipulation} attack. It also reduces gas costs since calling functions in embedded libraries requires less gas than external calls.
	
\subsubsection{Function visibility} We carefully define the visibility of each function. Most of the functions are declared as \texttt{External} (\eg \texttt{Approve()}, \texttt{Transfer()}, \etc) per specifications of \erc standard.

\subsection{Best practices and enhancements}
We also take into account a number of best practices that have been accepted by the Ethereum community to proactively prevent known vulnerabilities~\cite{TokenBP}. Again, we highlight several of these while placing the background details in the appendix.

\subsubsection{Compliance with \erc}
We implement all \erc functions to make it fully compatible with the standard. Compliance is important for ensuring that other DApps and web apps (\ie crypto-wallets, crypto-exchanges, web services, \etc) compose with \sys as expected. 

\subsubsection{External visibility}
To improve performance, we apply an \texttt{external} visibility (instead of \texttt{public} visibility in the standard) for interactive functions (\eg \texttt{approve()} and \texttt{transfer()}, \etc).  External functions can read arguments directly from non-persistent \texttt{calldata} instead of allocating persistent memory by the EVM. 

\subsubsection{Fail-Safe Mode}
We implement a `cease trade' operation that will freeze the token in the case of new security threats or new legal requirements (\eg Liberty Reserve ~\cite{LibertyReserve} or TON cryptocurrency~\cite{TON}). To freeze all functionality of \sys, the owner (or multiple parties) can call the function \texttt{pause()} which sets a lock variable. All critical methods are either marked with a \texttt{notPaused} modifier (in Solidity) or explicit check (in Vyper), that will throw exceptions until functionality is restored using \texttt{unpause()}. 

\subsubsection{Firing events}
We define nine extra events: \texttt{Buy}, \texttt{Sell}, \texttt{Received}, \texttt{Withdrawal}, \texttt{Pause}, \texttt{Change}, \texttt{ChangeOwner}, \texttt{Mint} and \texttt{Burn}. The name of each event indicates its function except \texttt{Change} event which logs any state variable updates. It can be used to watch for token inconsistent behavior (\eg via TokenScope~\cite{TokenScope}) and react accordingly. 
	
\subsubsection{Proxy contracts}
We choose to make \sys non-upgradable so it can be audited, and upgrades will not introduce new vulnerabilities that did not exist at the time of the initial audit. 
	
\subsubsection{Other enhancements}
We also follow other best practices such as not using batch processing in \texttt{sell()} function to avoid \textit{DoS with unexpected revert} issue, not using miner controlled variable in conditional statements, and not using \texttt{SELFDESTRUCT}.

\subsection{Implementing in Vyper vs. Solidity}
Although Vyper offers less features than Solidity (\eg no class inheritance, modifiers, inline assembly, function/operator overloading, \etc~\cite{SolidityDoc}), the Vyper compiler includes built-in security checks. Table~\ref{tab:compare} provides a comparison between the two from the perspective of \sys (see~\cite{Vyper1} for a broader comparison on vulnerabilities). Security and performance are advantages of Vyper. However, Vyper may not be a preferred option for production (``Vyper is beta software, use with care''~\cite{VyperReadme}), most of the auditing tools only support Solidity,\footnote{Vyper support is recently added to some tools (\eg Crytic-compile, Manticore and Echidna). Slither integration is still in progress~\cite{Crytic}} and Solidity currently enjoys widespread  implementation, developer tools, and developer experience.


\begin{table*}[]
	\centering
\centering
\scalebox{0.64}{
	\begin{tabular}{|l|l|c|c|l|}
		\hline
		\multicolumn{2}{|c|}{\multirow{2}{*}{\textbf{\begin{tabular}[c]{@{}c@{}}Vulnerability (Vul.) or \\ Best Practice (BP.)\end{tabular}}}} & \multicolumn{2}{c|}{\textbf{\begin{tabular}[c]{@{}c@{}}\sys\\ Implementation\end{tabular}}} & \multicolumn{1}{c|}{\multirow{2}{*}{\textbf{Comment}}} \\ \cline{3-4}
		\multicolumn{2}{|c|}{} & \textbf{Vyper} & \textbf{Solidity} & \multicolumn{1}{c|}{} \\ \hline
		Arithmetic Over/Under Flows & Vul. & + &  & \begin{tabular}[c]{@{}l@{}}- Vyper includes built-in checks for over/under flows.\\ - \texttt{SafeMath} library is required in Solidity to mitigate the attack.\end{tabular} \\ \hline
		Re-Entrancy & Vul. & + &  & \begin{tabular}[c]{@{}l@{}}- \texttt{@nonreentrant} decorator places a lock on functions to mitigate the attack.\\ - \texttt{noReentrancy} modifier is required in Solidity.\end{tabular} \\ \hline
		Unchecked return values & Vul. & + &  & \begin{tabular}[c]{@{}l@{}}- It is already addressed in Vyper.\\ - There is a need in Solidity to check return values explicitly.\end{tabular} \\ \hline
		Code readability & BP. & + &  & \begin{tabular}[c]{@{}l@{}}- No inheritance in Vyper enforces simpler design.\\ - Solidity allows inline assemblies which is riskier and decreases readability.\end{tabular} \\ \hline
		Contract complexity & BP. & + &  & - 300 lines in Vyper have the same functionality as the Solidity with 500 lines. \\ \hline
		Auditable & BP. &  & + & - Most of the auditing tools are able to analyze Solidity contracts. \\ \hline
		Compatibility & BP. &  & + & \begin{tabular}[c]{@{}l@{}}- Majority of the current Ethereum projects are based on Solidity.\\ - Developers are more familiar with Solidity than Vyper.\end{tabular} \\ \hline
		Production readiness & BP. &  & + & \begin{tabular}[c]{@{}l@{}}- Vyper is not as mature as Solidity in terms of stability, documentation, etc.\\ - Solidity is adapted by a larger development community.\end{tabular} \\ \hline
	\end{tabular}}
	\caption{Comparison of \sys implementation in Vyper and Solidity. The plus sign can be considered as an advantage. However, both versions of \sys offer the same level of security.}
	\label{tab:compare}
\end{table*}
	
\subsection{Need for another reference implementation}
The authors of the \erc standard reference two sample Solidity implementations: one that is actively maintained by OpenZeppelin~\cite{OpenZepplin} and one that has been deprecated by ConsenSys~\cite{ConsensysToken} (and now refers to the OpenZeppelin implementation). As expected, the OpenZeppelin template is very popular within the Solidity developers~\cite{OPZ1,OPZ2,OPZ3}.

OpenZeppelin's implementation is actually part of a small portfolio of implementations (ERC20, ERC721, ERC777, and ERC1155). Code reuse across the four implementations adds complexity for a developer that only wants \erc. This might be the reason for not supporting Vyper in OpenZeppelin's implementation. No inheritance in Vyper requires different implementation than the current object-oriented OpenZeppelin contracts. Further, most audit tools are not able to import libraries/interfaces from external files (\eg SafeMath.sol, IERC20.sol). By contrast, \sys uses a flat layout in a single file that is specific to \erc. It does not use inheritance in Solidity which allows similar implementation in Vyper.  


\sys makes other improvements over the OpenZeppelin implementation. For example, OpenZeppelin introduces two new functions to mitigate the multiple withdraw attack: \texttt{increaseAllowance()} and \texttt{decreaseAllowance()}. However these are not part of the \erc standard and are not  interoperable with other applications that expect to use \texttt{approve()} and \texttt{transferFrom()}. \sys secures \texttt{transferFrom()} to prevent the attack (following~\cite{ERC20MWA}) and is interoperable with legacy DApps and web apps. Additionally, \sys mitigates the \textit{frozen Ether} issue by introducing a \texttt{withdraw()} function, while ETH forced into the OpenZeppelin implementation is forever unrecoverable. Both contracts implement a \textit{fail-safe mode}, however this logic is internal to \sys, while OpenZeppelin requires an external \texttt{Pausable.sol} contract.

Diversity in software is important for robustness and security~\cite{FSA97,FHS97}. For \erc, a variety of implementations will reduce the impact of a single bug in a single implementation. For example, between 17 March 2017 and 13 July 2017, OpenZeppelin's implementation used the wrong interface and affected 130 tokens~\cite{ErcBug}. \sys increases the diversity of \erc Solidity implementations and addresses the lack of a reference implementation in Vyper.

\newcommand{\BP}{\texttt{BP}}
\newcommand{\fp}{$\oplus$}
\newcommand{\fail}{$\times$}
\newcommand{\noSWC}{$\bigcirc$}
\newcommand{\info}{$!$}
\newcommand{\pass}{$\checkmark$}
\newcommand{\na}{}
\newcommand{\tx}[1]{\fontfamily{cmss}\selectfont{\textbf{#1}}}
\newcommand{\ccl}{\cellcolor[HTML]{EFEFEF}}
\newcommand{\rcl}{\rowcolor[HTML]{EFEFEF}}
\newcolumntype{P}[1]{>{\centering\arraybackslash}p{#1}}

\begin{table*}
\centering
\begin{adjustbox}{max height=10cm}
\begin{tabular}{|P{3mm}|P{7mm}|m{105mm}|P{3mm}|P{3mm}|P{3mm}|P{3mm}|P{3mm}|P{3mm}|P{3mm}|}

\multicolumn{3}{c}{\.} &
\headrow{EY Token Review} &
\headrow{Smart Check} &
\headrow{Securify} &
\headrow{MythX (Mythril)} &
\headrow{Contract Guard} &
\headrow{Slither} &
\headrow{Odin} \\ \hline

\rcl\ccl & \ccl & \tx{Vulnerability or best practice} & \multicolumn{7}{c|}{\ccl} \\ \cline{3-3}
\rcl\multirow{-2}{*}{\ccl\tx{ID}} & \multirow{-2}{*}{\ccl\tx{SWC}} & Mitigation or recommendation & \multicolumn{7}{c|}{\multirow{-2}{*}{\ccl\tx{Security tools}}} \\ \hline

\multirow{2}{*}{1} & \multirow{2}{*}{100} & \tx{Function default visibility} & \multirow{2}{*}{\na} & \multirow{2}{*}{\pass} & \multirow{2}{*}{\na} & \multirow{2}{*}{\pass} & \multirow{2}{*}{\pass} & \multirow{2}{*}{\na} & \multirow{2}{*}{\pass} \\ \cline{3-3} & & Specifying function visibility, external, public, internal or private & & & & & & & \\ \hline
\multirow{2}{*}{2} & \multirow{2}{*}{101} & \tx{Integer Overflow and Underflow} & \multirow{2}{*}{\fp} & \multirow{2}{*}{\info} & \multirow{2}{*}{\na} & \multirow{2}{*}{\pass} & \multirow{2}{*}{\pass} & \multirow{2}{*}{\na} & \multirow{2}{*}{\pass} \\ \cline{3-3}
& & Utilizing the SafeMath library to mitigate over/under value assignments & & & & & & & \\ \hline
\multirow{2}{*}{3} & \multirow{2}{*}{102} & \tx{Outdated Compiler Version} & \multirow{2}{*}{\pass} & \multirow{2}{*}{\pass} & \multirow{2}{*}{\pass} & \multirow{2}{*}{\pass} & \multirow{2}{*}{\pass} & \multirow{2}{*}{\pass} & \multirow{2}{*}{\fail} \\ \cline{3-3}
& & Using proper Solidity version to protect against compiler attacks & & & & & & & \\ \hline
\multirow{2}{*}{4} & \multirow{2}{*}{103} & \tx{Floating Pragma} & \multirow{2}{*}{\na} & \multirow{2}{*}{\pass} & \multirow{2}{*}{\pass} & \multirow{2}{*}{\pass} & \multirow{2}{*}{\na} & \multirow{2}{*}{\pass} & \multirow{2}{*}{\pass} \\ \cline{3-3}
& & Locking the pragma to avoid deployments using outdated compiler version & & & & & & & \\ \hline
\multirow{2}{*}{5} & \multirow{2}{*}{104} & \tx{Unchecked Call Return Value} & \multirow{2}{*}{\fp} & \multirow{2}{*}{\na} & \multirow{2}{*}{\pass} & \multirow{2}{*}{\pass} & \multirow{2}{*}{\pass} & \multirow{2}{*}{\fp} & \multirow{2}{*}{\pass} \\ \cline{3-3}
& & Checking call() return value to prevent unexpected behavior in DApps & & & & & & & \\ \hline
\multirow{2}{*}{6} & \multirow{2}{*}{105} & \tx{Unprotected Ether Withdrawal} & \multirow{2}{*}{\na} & \multirow{2}{*}{\info} & \multirow{2}{*}{\na} & \multirow{2}{*}{\pass} & \multirow{2}{*}{\na} & \multirow{2}{*}{\pass} & \multirow{2}{*}{\pass} \\ \cline{3-3}
& & Authorizing only trusted parties to trigger ETH withdrawals & & & & & & & \\ \hline
\multirow{2}{*}{7} & \multirow{2}{*}{106} & \tx{Unprotected SELFDESTRUCT Instruction} & \multirow{2}{*}{\na} & \multirow{2}{*}{\na} & \multirow{2}{*}{\pass} & \multirow{2}{*}{\pass} & \multirow{2}{*}{\na} & \multirow{2}{*}{\pass} & \multirow{2}{*}{\pass} \\ \cline{3-3}
& & Removing self-destruct functionality or approving it by multiple parties & & & & & & & \\ \hline
\multirow{2}{*}{8} & \multirow{2}{*}{107} & \tx{Re-entrancy} & \multirow{2}{*}{\na} & \multirow{2}{*}{\pass} & \multirow{2}{*}{\fp} & \multirow{2}{*}{\fp} & \multirow{2}{*}{\fp} & \multirow{2}{*}{\pass} & \multirow{2}{*}{\pass} \\ \cline{3-3}
& & Using CEI and Mutex to mitigate self-function and cross-function attack & & & & & & & \\ \hline
\multirow{2}{*}{9} & \multirow{2}{*}{108} & \tx{State variable default visibility} & \multirow{2}{*}{\pass} & \multirow{2}{*}{\pass} & \multirow{2}{*}{\pass} & \multirow{2}{*}{\pass} & \multirow{2}{*}{\pass} & \multirow{2}{*}{\na} & \multirow{2}{*}{\pass} \\ \cline{3-3}
& & Specifying visibility of all variables, public, private or internal & & & & & & & \\ \hline
\multirow{2}{*}{10} & \multirow{2}{*}{109} & \tx{Uninitialized Storage Pointer} & \multirow{2}{*}{\pass} & \multirow{2}{*}{\pass} & \multirow{2}{*}{\pass} & \multirow{2}{*}{\pass} & \multirow{2}{*}{\pass} & \multirow{2}{*}{\pass} & \multirow{2}{*}{\pass} \\ \cline{3-3}
& & Initializing variables upon declaration to prevent unexpected storage access & & & & & & & \\ \hline
\multirow{2}{*}{11} & \multirow{2}{*}{110} & \tx{Assert Violation} & \multirow{2}{*}{\na} & \multirow{2}{*}{\pass} & \multirow{2}{*}{\na} & \multirow{2}{*}{\pass} & \multirow{2}{*}{\na} & \multirow{2}{*}{\na} & \multirow{2}{*}{\pass} \\ \cline{3-3}
& & Using require() statement to validate inputs, checking efficiency of the code & & & & & & & \\ \hline
\multirow{2}{*}{12} & \multirow{2}{*}{111} & \tx{Use of Deprecated Solidity Functions} & \multirow{2}{*}{\na} & \multirow{2}{*}{\pass} & \multirow{2}{*}{\na} & \multirow{2}{*}{\pass} & \multirow{2}{*}{\pass} & \multirow{2}{*}{\pass} & \multirow{2}{*}{\pass} \\ \cline{3-3}
& & Using new alternatives functions such as keccak256() instead of sha3() & & & & & & & \\ \hline
\multirow{2}{*}{13} & \multirow{2}{*}{112} & \tx{Delegatecall to untrusted callee} & \multirow{2}{*}{\na} & \multirow{2}{*}{\fp} & \multirow{2}{*}{\fp} & \multirow{2}{*}{\pass} & \multirow{2}{*}{\pass} & \multirow{2}{*}{\pass} & \multirow{2}{*}{\pass} \\ \cline{3-3}
& & Calling into trusted contracts to avoid storage access by malicious contracts & & & & & & & \\ \hline
\multirow{2}{*}{14} & \multirow{2}{*}{113} & \tx{DoS with Failed Call} & \multirow{2}{*}{\pass} & \multirow{2}{*}{\pass} & \multirow{2}{*}{\na} & \multirow{2}{*}{\pass} & \multirow{2}{*}{\pass} & \multirow{2}{*}{\na} & \multirow{2}{*}{\pass} \\ \cline{3-3}
& & Avoid multiple external calls where one error may fail other transactions & & & & & & & \\ \hline
\multirow{2}{*}{15} & \multirow{2}{*}{114} & \tx{Transaction Order Dependence} & \multirow{2}{*}{\fp} & \multirow{2}{*}{\na} & \multirow{2}{*}{\pass} & \multirow{2}{*}{\pass} & \multirow{2}{*}{\na} & \multirow{2}{*}{\na} & \multirow{2}{*}{\pass} \\ \cline{3-3}
& & Preventing race conditions by securing approve() or transferFrom() & & & & & & & \\ \hline
\multirow{2}{*}{16} & \multirow{2}{*}{115} & \tx{Authorization through tx.origin} & \multirow{2}{*}{\pass} & \multirow{2}{*}{\pass} & \multirow{2}{*}{\pass} & \multirow{2}{*}{\pass} & \multirow{2}{*}{\pass} & \multirow{2}{*}{\pass} & \multirow{2}{*}{\pass} \\ \cline{3-3}
& & Using msg.sender to authorize transaction initiator instead of originator & & & & & & & \\ \hline
\multirow{2}{*}{17} & \multirow{2}{*}{116} & \tx{Block values as a proxy for time} & \multirow{2}{*}{\pass} & \multirow{2}{*}{\pass} & \multirow{2}{*}{\pass} & \multirow{2}{*}{\pass} & \multirow{2}{*}{\pass} & \multirow{2}{*}{\na} & \multirow{2}{*}{\pass} \\ \cline{3-3}
& & Not using block.timestamp or block.number to perform functionalities & & & & & & & \\ \hline
\multirow{2}{*}{18} & \multirow{2}{*}{117} & \tx{Signature Malleability} & \multirow{2}{*}{\na} & \multirow{2}{*}{\na} & \multirow{2}{*}{\na} & \multirow{2}{*}{\pass} & \multirow{2}{*}{\na} & \multirow{2}{*}{\na} & \multirow{2}{*}{\pass} \\ \cline{3-3}
& & Not using signed message hash to avoid signatures alteration & & & & & & & \\ \hline
\multirow{2}{*}{19} & \multirow{2}{*}{118} & \tx{Incorrect Constructor Name} & \multirow{2}{*}{\na} & \multirow{2}{*}{\pass} & \multirow{2}{*}{\na} & \multirow{2}{*}{\pass} & \multirow{2}{*}{\na} & \multirow{2}{*}{\na} & \multirow{2}{*}{\pass} \\ \cline{3-3}
& & Using constructor keyword which does not match with contract name & & & & & & & \\ \hline
\multirow{2}{*}{20} & \multirow{2}{*}{119} & \tx{Shadowing State Variables} & \multirow{2}{*}{\na} & \multirow{2}{*}{\na} & \multirow{2}{*}{\pass} & \multirow{2}{*}{\pass} & \multirow{2}{*}{\pass} & \multirow{2}{*}{\pass} & \multirow{2}{*}{\pass} \\ \cline{3-3}
& & Removing any variable ambiguities when inheriting other contracts & & & & & & & \\ \hline
\multirow{2}{*}{21} & \multirow{2}{*}{120} & \tx{Weak Sources of Randomness from Chain Attributes} & \multirow{2}{*}{\pass} & \multirow{2}{*}{\pass} & \multirow{2}{*}{\na} & \multirow{2}{*}{\pass} & \multirow{2}{*}{\pass} & \multirow{2}{*}{\na} & \multirow{2}{*}{\pass} \\ \cline{3-3}
& & Using oracles as source of randomness instead of block.timestamp & & & & & & & \\ \hline
\multirow{2}{*}{22} & \multirow{2}{*}{121} & \tx{Missing Protection against Signature Replay Attacks} & \multirow{2}{*}{\na} & \multirow{2}{*}{\na} & \multirow{2}{*}{\na} & \multirow{2}{*}{\pass} & \multirow{2}{*}{\na} & \multirow{2}{*}{\na} & \multirow{2}{*}{\pass} \\ \cline{3-3}
& & Storing every message hash to perform signature verification & & & & & & & \\ \hline
\multirow{2}{*}{23} & \multirow{2}{*}{122} & \tx{Lack of Proper Signature Verification} & \multirow{2}{*}{\na} & \multirow{2}{*}{\na} & \multirow{2}{*}{\na} & \multirow{2}{*}{\pass} & \multirow{2}{*}{\na} & \multirow{2}{*}{\na} & \multirow{2}{*}{\pass} \\ \cline{3-3}
& & Using alternate verification schemes if allowing off-chain signing & & & & & & & \\ \hline
\multirow{2}{*}{24} & \multirow{2}{*}{123} & \tx{Requirement Violation} & \multirow{2}{*}{\na} & \multirow{2}{*}{\pass} & \multirow{2}{*}{\pass} & \multirow{2}{*}{\pass} & \multirow{2}{*}{\na} & \multirow{2}{*}{\na} & \multirow{2}{*}{\pass} \\ \cline{3-3}
& & Checking the code for allowing only valid external inputs & & & & & & & \\ \hline
\multirow{2}{*}{25} & \multirow{2}{*}{124} & \tx{Write to Arbitrary Storage Location} & \multirow{2}{*}{\na} & \multirow{2}{*}{\pass} & \multirow{2}{*}{\pass} & \multirow{2}{*}{\pass} & \multirow{2}{*}{\na} & \multirow{2}{*}{\na} & \multirow{2}{*}{\pass} \\ \cline{3-3}
& & Controlling write to storage to prevent storage corruption by attackers & & & & & & & \\ \hline
\multirow{2}{*}{26} & \multirow{2}{*}{125} & \tx{Incorrect Inheritance Order} & \multirow{2}{*}{\na} & \multirow{2}{*}{\na} & \multirow{2}{*}{\na} & \multirow{2}{*}{\pass} & \multirow{2}{*}{\na} & \multirow{2}{*}{\na} & \multirow{2}{*}{\pass} \\ \cline{3-3}
& & Inheriting from more general to specific when there are identical functions & & & & & & & \\ \hline
\multirow{2}{*}{27} & \multirow{2}{*}{126} & \tx{Insufficient Gas Griefing} & \multirow{2}{*}{\na} & \multirow{2}{*}{\pass} & \multirow{2}{*}{\na} & \multirow{2}{*}{\na} & \multirow{2}{*}{\na} & \multirow{2}{*}{\na} & \multirow{2}{*}{\pass} \\ \cline{3-3}
& & Allowing trusted forwarders to relay transactions & & & & & & & \\ \hline

\end{tabular}
\end{adjustbox}	
\caption{Auditing results of 7 smart contract analysis tools on \sys. \pass=Passed audit, \fp=False positive, \fail=Failed audit, Empty=Not supported audit by the tool, \info=Informational, \noSWC=Tool specific audit (No SWC registry), BP=Best practice\label{tab:result1}}
\end{table*}


\begin{table*}
\centering
\begin{adjustbox}{max height=10cm}
\begin{tabular}{|P{3mm}|P{7mm}|m{105mm}|P{3mm}|P{3mm}|P{3mm}|P{3mm}|P{3mm}|P{3mm}|P{3mm}|}
	
\multicolumn{3}{c}{\.} &
\headrow{EY Token Review} &
\headrow{Smart Check} &
\headrow{Securify} &
\headrow{MythX (Mythril)} &
\headrow{Contract Guard} &
\headrow{Slither} &
\headrow{Odin} \\ \hline

\rcl\ccl & \ccl & \tx{Vulnerability or best practice} & \multicolumn{7}{c|}{\ccl} \\ \cline{3-3}
\rcl\multirow{-2}{*}{\ccl\tx{ID}} & \multirow{-2}{*}{\ccl\tx{SWC}} & Mitigation or recommendation & \multicolumn{7}{c|}{\multirow{-2}{*}{\ccl\tx{Security tools}}} \\ \hline

\multirow{2}{*}{28} & \multirow{2}{*}{127} & \tx{Arbitrary Jump with Function Type Variable} & \multirow{2}{*}{\na} & \multirow{2}{*}{\pass} & \multirow{2}{*}{\pass} & \multirow{2}{*}{\pass} & \multirow{2}{*}{\na} & \multirow{2}{*}{\pass} & \multirow{2}{*}{\pass} \\ \cline{3-3}
& & Minimizing use of assembly in the code & & & & & & & \\ \hline
\multirow{2}{*}{29} & \multirow{2}{*}{128} & \tx{DoS With Block Gas Limit} & \multirow{2}{*}{\pass} & \multirow{2}{*}{\pass} & \multirow{2}{*}{\pass} & \multirow{2}{*}{\pass} & \multirow{2}{*}{\pass} & \multirow{2}{*}{\pass} & \multirow{2}{*}{\pass} \\ \cline{3-3}
& & Avoiding loops across the code that may consume considerable resources & & & & & & & \\ \hline
\multirow{2}{*}{30} & \multirow{2}{*}{129} & \tx{Typographical Error} & \multirow{2}{*}{\na} & \multirow{2}{*}{\na} & \multirow{2}{*}{\na} & \multirow{2}{*}{\pass} & \multirow{2}{*}{\na} & \multirow{2}{*}{\na} & \multirow{2}{*}{\pass} \\ \cline{3-3}
& & Using SafeMath library or performing checks on any math operation & & & & & & & \\ \hline
\multirow{2}{*}{31} & \multirow{2}{*}{130} & \tx{Right-To-Left-Override control character (U+202E)} & \multirow{2}{*}{\na} & \multirow{2}{*}{\na} & \multirow{2}{*}{\pass} & \multirow{2}{*}{\pass} & \multirow{2}{*}{\pass} & \multirow{2}{*}{\pass} & \multirow{2}{*}{\pass} \\ \cline{3-3}
& & Avoiding U+202E character which forces RTL text rendering & & & & & & & \\ \hline
\multirow{2}{*}{32} & \multirow{2}{*}{131} & \tx{Presence of unused variables} & \multirow{2}{*}{\na} & \multirow{2}{*}{\pass} & \multirow{2}{*}{\pass} & \multirow{2}{*}{\na} & \multirow{2}{*}{\pass} & \multirow{2}{*}{\pass} & \multirow{2}{*}{\fp} \\ \cline{3-3} & & Removing all unused variables to decrease gas consumption & & & & & & & \\ \hline
\multirow{2}{*}{33} & \multirow{2}{*}{132} & \tx{Unexpected Ether balance} & \multirow{2}{*}{\na} & \multirow{2}{*}{\pass} & \multirow{2}{*}{\pass} & \multirow{2}{*}{\na} & \multirow{2}{*}{\pass} & \multirow{2}{*}{\pass} & \multirow{2}{*}{\pass} \\ \cline{3-3} & & Avoiding Ether balance check in the code (\eg this.balance == 0.24 Ether) & & & & & & & \\ \hline
\multirow{2}{*}{34} & \multirow{2}{*}{133} & \tx{Hash Collisions With Variable Length Arguments} & \multirow{2}{*}{\na} & \multirow{2}{*}{\na} & \multirow{2}{*}{\na} & \multirow{2}{*}{\na} & \multirow{2}{*}{\na} & \multirow{2}{*}{\na} & \multirow{2}{*}{\pass} \\ \cline{3-3} & & Using abi.encode() instead of abi.encodePacked() to prevent hash collision & & & & & & & \\ \hline
\multirow{2}{*}{35} & \multirow{2}{*}{134} & \tx{Message call with hardcoded gas amount} & \multirow{2}{*}{\na} & \multirow{2}{*}{\fp} & \multirow{2}{*}{\fp} & \multirow{2}{*}{\pass} & \multirow{2}{*}{\pass} & \multirow{2}{*}{\na} & \multirow{2}{*}{\pass} \\ \cline{3-3} & & Using .call.value()("") which is compatible with EIP1884 & & & & & & & \\ \hline
\multirow{2}{*}{36} & \multirow{2}{*}{135} & \tx{Code With No Effects} & \multirow{2}{*}{\na} & \multirow{2}{*}{\pass} & \multirow{2}{*}{\na} & \multirow{2}{*}{\na} & \multirow{2}{*}{\na} & \multirow{2}{*}{\na} & \multirow{2}{*}{\pass} \\ \cline{3-3} & & Writing unit tests to ensure producing the intended effects by DApps & & & & & & & \\ \hline
\multirow{2}{*}{37} & \multirow{2}{*}{136} & \tx{Unencrypted Private Data On-Chain} & \multirow{2}{*}{\na} & \multirow{2}{*}{\info} & \multirow{2}{*}{\na} & \multirow{2}{*}{\na} & \multirow{2}{*}{\na} & \multirow{2}{*}{\na} & \multirow{2}{*}{\pass} \\ \cline{3-3} & & Storing un-encrypted private data off-chain & & & & & & & \\ \hline
\multirow{2}{*}{38} & \multirow{2}{*}{\noSWC} & \tx{Allowance decreases upon transfer} & \multirow{2}{*}{\pass} & \multirow{2}{*}{\na} & \multirow{2}{*}{\na} & \multirow{2}{*}{\na} & \multirow{2}{*}{\na} & \multirow{2}{*}{\na} & \multirow{2}{*}{\na} \\ \cline{3-3} & & Decreasing allowance in transferFrom() method & & & & & & & \\ \hline
\multirow{2}{*}{39} & \multirow{2}{*}{\noSWC} & \tx{Allowance function returns an accurate value} & \multirow{2}{*}{\pass} & \multirow{2}{*}{\na} & \multirow{2}{*}{\na} & \multirow{2}{*}{\na} & \multirow{2}{*}{\na} & \multirow{2}{*}{\na} & \multirow{2}{*}{\na} \\ \cline{3-3} & & Returning only value from the mapping instead of internal function logic & & & & & & & \\ \hline
\multirow{2}{*}{40} & \multirow{2}{*}{\noSWC} & \tx{It is possible to cancel an existing allowance} & \multirow{2}{*}{\pass} & \multirow{2}{*}{\pass} & \multirow{2}{*}{\na} & \multirow{2}{*}{\na} & \multirow{2}{*}{\na} & \multirow{2}{*}{\na} & \multirow{2}{*}{\na} \\ \cline{3-3} & & Possibility of setting allowance to 0 to revoke previous allowances & & & & & & & \\ \hline
\multirow{2}{*}{41} & \multirow{2}{*}{\noSWC} & \tx{A transfer with an insufficient amount is reverted} & \multirow{2}{*}{\pass} & \multirow{2}{*}{\na} & \multirow{2}{*}{\na} & \multirow{2}{*}{\na} & \multirow{2}{*}{\na} & \multirow{2}{*}{\pass} & \multirow{2}{*}{\na} \\ \cline{3-3} & & Checking balances in transfer() method before updating balances & & & & & & & \\ \hline
\multirow{2}{*}{42} & \multirow{2}{*}{\noSWC} & \tx{Upon sending funds, the sender's balance is updated} & \multirow{2}{*}{\pass} & \multirow{2}{*}{\na} & \multirow{2}{*}{\na} & \multirow{2}{*}{\na} & \multirow{2}{*}{\na} & \multirow{2}{*}{\na} & \multirow{2}{*}{\na} \\ \cline{3-3} & & Updating balances in transfer() or transferFrom() methods & & & & & & & \\ \hline
\multirow{2}{*}{43} & \multirow{2}{*}{\noSWC} & \tx{The Transfer event correctly logged} & \multirow{2}{*}{\pass} & \multirow{2}{*}{\na} & \multirow{2}{*}{\na} & \multirow{2}{*}{\na} & \multirow{2}{*}{\na} & \multirow{2}{*}{\na} & \multirow{2}{*}{\na} \\ \cline{3-3} & & Emitting Transfer event in transfer() or transferFrom() functions & & & & & & & \\ \hline
\multirow{2}{*}{44} & \multirow{2}{*}{\noSWC} & \tx{Transfer an amount that is greater than the allowance} & \multirow{2}{*}{\pass} & \multirow{2}{*}{\na} & \multirow{2}{*}{\na} & \multirow{2}{*}{\na} & \multirow{2}{*}{\na} & \multirow{2}{*}{\na} & \multirow{2}{*}{\na} \\ \cline{3-3} & & Checking balances in transferFrom() method before updating balances & & & & & & & \\ \hline
\multirow{2}{*}{45} & \multirow{2}{*}{\noSWC} & \tx{Risk of short address attack is minimized} & \multirow{2}{*}{\pass} & \multirow{2}{*}{\na} & \multirow{2}{*}{\na} & \multirow{2}{*}{\na} & \multirow{2}{*}{\pass} & \multirow{2}{*}{\na} & \multirow{2}{*}{\na} \\ \cline{3-3} & & Using recent Solidity version to mitigate the attack & & & & & & & \\ \hline
\multirow{2}{*}{46} & \multirow{2}{*}{\noSWC} & \tx{Function names are unique} & \multirow{2}{*}{\pass} & \multirow{2}{*}{\na} & \multirow{2}{*}{\na} & \multirow{2}{*}{\na} & \multirow{2}{*}{\na} & \multirow{2}{*}{\pass} & \multirow{2}{*}{\na} \\ \cline{3-3} & & No function overloading to avoid unexpected behavior & & & & & & & \\ \hline
\multirow{2}{*}{47} & \multirow{2}{*}{\noSWC} & \tx{Using miner controlled variables} & \multirow{2}{*}{\pass} & \multirow{2}{*}{\pass} & \multirow{2}{*}{\pass} & \multirow{2}{*}{\pass} & \multirow{2}{*}{\pass} & \multirow{2}{*}{\pass} & \multirow{2}{*}{\na} \\ \cline{3-3} & & Avoiding block.number, block.timestamp, block.difficulty, now, etc & & & & & & & \\ \hline
\multirow{2}{*}{48} & \multirow{2}{*}{\noSWC} & \tx{Use of return in constructor} & \multirow{2}{*}{\na} & \multirow{2}{*}{\pass} & \multirow{2}{*}{\na} & \multirow{2}{*}{\na} & \multirow{2}{*}{\na} & \multirow{2}{*}{\na} & \multirow{2}{*}{\na} \\ \cline{3-3} & & Not using return in contract's constructor & & & & & & & \\ \hline
\multirow{2}{*}{49} & \multirow{2}{*}{\noSWC} & \tx{Throwing exceptions in transfer() and transferFrom()} & \multirow{2}{*}{\na} & \multirow{2}{*}{\pass} & \multirow{2}{*}{\na} & \multirow{2}{*}{\na} & \multirow{2}{*}{\na} & \multirow{2}{*}{\pass} & \multirow{2}{*}{\na} \\ \cline{3-3} & & Returning true after successful execution or raising exception in failures & & & & & & & \\ \hline
\multirow{2}{*}{50} & \multirow{2}{*}{\noSWC} & \tx{State variables that could be declared constant} & \multirow{2}{*}{\na} & \multirow{2}{*}{\na} & \multirow{2}{*}{\na} & \multirow{2}{*}{\na} & \multirow{2}{*}{\na} & \multirow{2}{*}{\pass} & \multirow{2}{*}{\na} \\ \cline{3-3} & & Adding constant attribute to variables like name, symbol, decimals, etc & & & & & & & \\ \hline
\multirow{2}{*}{51} & \multirow{2}{*}{\noSWC} & \tx{Tautology or contradiction} & \multirow{2}{*}{\na} & \multirow{2}{*}{\na} & \multirow{2}{*}{\na} & \multirow{2}{*}{\na} & \multirow{2}{*}{\na} & \multirow{2}{*}{\pass} & \multirow{2}{*}{\na} \\ \cline{3-3} & & Fixing comparison in the code that are always true or false & & & & & & & \\ \hline
\multirow{2}{*}{52} & \multirow{2}{*}{\noSWC} & \tx{Divide before multiply} & \multirow{2}{*}{\na} & \multirow{2}{*}{\na} & \multirow{2}{*}{\na} & \multirow{2}{*}{\na} & \multirow{2}{*}{\na} & \multirow{2}{*}{\pass} & \multirow{2}{*}{\na} \\ \cline{3-3} & & Ordering multiplication prior division to avoid integer truncation & & & & & & & \\ \hline
\multirow{2}{*}{53} & \multirow{2}{*}{\noSWC} & \tx{Unchecked Send} & \multirow{2}{*}{\na} & \multirow{2}{*}{\na} & \multirow{2}{*}{\na} & \multirow{2}{*}{\na} & \multirow{2}{*}{\na} & \multirow{2}{*}{\pass} & \multirow{2}{*}{\na} \\ \cline{3-3} & & Ensuring that the return value of send() is always checked & & & & & & & \\ \hline
\multirow{2}{*}{54} & \multirow{2}{*}{\BP} & \tx{Too many digits} & \multirow{2}{*}{\na} & \multirow{2}{*}{\na} & \multirow{2}{*}{\na} & \multirow{2}{*}{\na} & \multirow{2}{*}{\na} & \multirow{2}{*}{\pass} & \multirow{2}{*}{\na} \\ \cline{3-3} & & Using scientific notation to make the code readable and simpler to debug & & & & & & & \\ \hline

\end{tabular}
\end{adjustbox}	
\caption{Continuation of Table\ref{tab:result1}.\label{tab:result2}}
\end{table*}


\begin{table*}
\centering
\begin{adjustbox}{max height=10cm}
\begin{tabular}{|P{3mm}|P{7mm}|m{95mm}|P{7mm}|P{7mm}|P{7mm}|P{7mm}|P{7mm}|P{7mm}|P{7mm}|}

\multicolumn{3}{c}{\.} &
\headrow{EY Token Review} &
\headrow{Smart Check} &
\headrow{Securify} &
\headrow{MythX (Mythril)} &
\headrow{Contract Guard} &
\headrow{Slither} &
\headrow{Odin} \\ \hline

\rcl\ccl & \ccl & \tx{Vulnerability or best practice} & \multicolumn{7}{c|}{\ccl} \\ \cline{3-3}
\rcl\multirow{-2}{*}{\ccl\tx{ID}} & \multirow{-2}{*}{\ccl\tx{SWC}} & Mitigation or recommendation & \multicolumn{7}{c|}{\multirow{-2}{*}{\ccl\tx{Security tools}}} \\ \hline

\multirow{2}{*}{55} & \multirow{2}{*}{\BP} & \tx{The decreaseAllowance definition follows the standard} & \multirow{2}{*}{\pass} & \multirow{2}{*}{\na} & \multirow{2}{*}{\na} & \multirow{2}{*}{\na} & \multirow{2}{*}{\na} & \multirow{2}{*}{\na} & \multirow{2}{*}{\na} \\ \cline{3-3} & & Defining decreaseAllowance input and output variables as standard & & & & & & & \\ \hline
\multirow{2}{*}{56} & \multirow{2}{*}{\BP} & \tx{The increaseAllowance definition follows the standard} & \multirow{2}{*}{\pass} & \multirow{2}{*}{\na} & \multirow{2}{*}{\na} & \multirow{2}{*}{\na} & \multirow{2}{*}{\na} & \multirow{2}{*}{\na} & \multirow{2}{*}{\na} \\ \cline{3-3} & & Defining increaseAllowance input and output variables as standard & & & & & & & \\ \hline
\multirow{2}{*}{57} & \multirow{2}{*}{\BP} & \tx{Minimize attack surface} & \multirow{2}{*}{\pass} & \multirow{2}{*}{\pass} & \multirow{2}{*}{\pass} & \multirow{2}{*}{\na} & \multirow{2}{*}{\na} & \multirow{2}{*}{\na} & \multirow{2}{*}{\na} \\ \cline{3-3} & & Checking whether all the external functions are necessary or not & & & & & & & \\ \hline
\multirow{2}{*}{58} & \multirow{2}{*}{\BP} & \tx{Transfer to the burn address is reverted} & \multirow{2}{*}{\pass} & \multirow{2}{*}{\na} & \multirow{2}{*}{\na} & \multirow{2}{*}{\na} & \multirow{2}{*}{\na} & \multirow{2}{*}{\na} & \multirow{2}{*}{\na} \\ \cline{3-3} & & Reverting transfer to 0x0 due to risk of total supply reduction & & & & & & & \\ \hline
\multirow{2}{*}{59} & \multirow{2}{*}{\BP} & \tx{Source code is decentralized} & \multirow{2}{*}{\pass} & \multirow{2}{*}{\pass} & \multirow{2}{*}{\na} & \multirow{2}{*}{\na} & \multirow{2}{*}{\na} & \multirow{2}{*}{\na} & \multirow{2}{*}{\na} \\ \cline{3-3} & & Not using hard-coded addresses in the code & & & & & & & \\ \hline
\multirow{2}{*}{60} & \multirow{2}{*}{\BP} & \tx{Funds can be held only by user-controlled wallets} & \multirow{2}{*}{\info} & \multirow{2}{*}{\na} & \multirow{2}{*}{\na} & \multirow{2}{*}{\na} & \multirow{2}{*}{\na} & \multirow{2}{*}{\na} & \multirow{2}{*}{\na} \\ \cline{3-3} & & Transferring tokens to users to avoid creating a secondary market & & & & & & & \\ \hline
\multirow{2}{*}{61} & \multirow{2}{*}{\BP} & \tx{Code logic is simple to understand} & \multirow{2}{*}{\pass} & \multirow{2}{*}{\pass} & \multirow{2}{*}{\na} & \multirow{2}{*}{\na} & \multirow{2}{*}{\na} & \multirow{2}{*}{\na} & \multirow{2}{*}{\na} \\ \cline{3-3} & & Avoiding code nesting which makes the code less intuitive & & & & & & & \\ \hline
\multirow{2}{*}{62} & \multirow{2}{*}{\BP} & \tx{All functions are documented} & \multirow{2}{*}{\pass} & \multirow{2}{*}{\na} & \multirow{2}{*}{\na} &\multirow{2}{*}{\na} & \multirow{2}{*}{\na} & \multirow{2}{*}{\na} & \multirow{2}{*}{\na} \\ \cline{3-3} & & Using NatSpec format to explain expected behavior of functions & & & & & & & \\ \hline
\multirow{2}{*}{63} & \multirow{2}{*}{\BP} & \tx{The Approval event is correctly logged} & \multirow{2}{*}{\pass} & \multirow{2}{*}{\na} & \multirow{2}{*}{\na} & \multirow{2}{*}{\na} & \multirow{2}{*}{\na} & \multirow{2}{*}{\na} & \multirow{2}{*}{\na} \\ \cline{3-3} & & Emitting Approval event in the approve() method & & & & & & & \\ \hline
\multirow{2}{*}{64} & \multirow{2}{*}{\BP} & \tx{Acceptable gas cost of the approve() function} & \multirow{2}{*}{\info} & \multirow{2}{*}{\na} & \multirow{2}{*}{\na} & \multirow{2}{*}{\na} & \multirow{2}{*}{\na} & \multirow{2}{*}{\na} & \multirow{2}{*}{\na} \\ \cline{3-3} & & Checking for maximum 50000 gas cost when executing the approve() & & & & & & & \\ \hline
\multirow{2}{*}{65} & \multirow{2}{*}{\BP} & \tx{Acceptable gas cost of the transfer() function} & \multirow{2}{*}{\info} & \multirow{2}{*}{\na} & \multirow{2}{*}{\na} & \multirow{2}{*}{\na} & \multirow{2}{*}{\na} & \multirow{2}{*}{\na} & \multirow{2}{*}{\na} \\ \cline{3-3} & & Checking for maximum 60000 gas cost when executing the transfer() & & & & & & & \\ \hline
\multirow{2}{*}{66} & \multirow{2}{*}{\BP} & \tx{Emitting event when state changes} & \multirow{2}{*}{\pass} & \multirow{2}{*}{\na} & \multirow{2}{*}{\na} & \multirow{2}{*}{\na} & \multirow{2}{*}{\na} & \multirow{2}{*}{\na} & \multirow{2}{*}{\na} \\ \cline{3-3} & & Emitting Change event when changing state variable values & & & & & & & \\ \hline
\multirow{2}{*}{67} & \multirow{2}{*}{\BP} & \tx{Use of unindexed arguments} & \multirow{2}{*}{\na} & \multirow{2}{*}{\pass} & \multirow{2}{*}{\na} & \multirow{2}{*}{\na} & \multirow{2}{*}{\pass} & \multirow{2}{*}{\pass} & \multirow{2}{*}{\na} \\ \cline{3-3} & & Using indexed arguments to facilitate external tools log searching & & & & & & & \\ \hline
\multirow{2}{*}{68} & \multirow{2}{*}{\BP} & \tx{\erc compliance} & \multirow{2}{*}{\pass} & \multirow{2}{*}{\pass} & \multirow{2}{*}{\pass} & \multirow{2}{*}{\na} & \multirow{2}{*}{\pass} & \multirow{2}{*}{\pass} & \multirow{2}{*}{\na} \\ \cline{3-3} & & Implementing all 6 functions and 2 events as specified in EIP-20 & & & & & & & \\ \hline
\multirow{2}{*}{69} & \multirow{2}{*}{\BP} & \tx{Conformance to naming conventions} & \multirow{2}{*}{\na} & \multirow{2}{*}{\na} & \multirow{2}{*}{\na} & \multirow{2}{*}{\na} & \multirow{2}{*}{\na} & \multirow{2}{*}{\pass} & \multirow{2}{*}{\na} \\ \cline{3-3} & & Following the Solidity naming convention to avoid confusion & & & & & & & \\ \hline
\multirow{2}{*}{70} & \multirow{2}{*}{\BP} & \tx{Token decimal} & \multirow{2}{*}{\pass} & \multirow{2}{*}{\na} & \multirow{2}{*}{\na} & \multirow{2}{*}{\na} & \multirow{2}{*}{\na} & \multirow{2}{*}{\na} & \multirow{2}{*}{\na} \\ \cline{3-3} & & Declaring token decimal for external apps when displaying balances & & & & & & & \\ \hline
\multirow{2}{*}{71} & \multirow{2}{*}{\BP} & \tx{Locked money (Freezing ETH)} & \multirow{2}{*}{\na} &\multirow{2}{*}{\pass} & \multirow{2}{*}{\na} & \multirow{2}{*}{\na} & \multirow{2}{*}{\pass} & \multirow{2}{*}{\pass} & \multirow{2}{*}{\na} \\ \cline{3-3} & & Implementing withdraw/reject functions to avoid ETH lost & & & & & & & \\ \hline
\multirow{2}{*}{72} & \multirow{2}{*}{\BP} & \tx{Malicious libraries} & \multirow{2}{*}{\na} & \multirow{2}{*}{\pass} & \multirow{2}{*}{\na} & \multirow{2}{*}{\na} & \multirow{2}{*}{\na} & \multirow{2}{*}{\na} & \multirow{2}{*}{\na} \\ \cline{3-3} & & Not using modifiable third-party libraries & & & & & & & \\ \hline
\multirow{2}{*}{73} & \multirow{2}{*}{\BP} & \tx{Payable fallback function} & \multirow{2}{*}{\na} & \multirow{2}{*}{\pass} & \multirow{2}{*}{\na} & \multirow{2}{*}{\na} & \multirow{2}{*}{\pass} & \multirow{2}{*}{\na} & \multirow{2}{*}{\na} \\ \cline{3-3} & & Adding either fallback() or receive() function to receive ETH & & & & & & & \\ \hline
\multirow{2}{*}{74} & \multirow{2}{*}{\BP} & \tx{Prefer external to public visibility level} & \multirow{2}{*}{\na} & \multirow{2}{*}{\pass} & \multirow{2}{*}{\na} & \multirow{2}{*}{\na} & \multirow{2}{*}{\na} & \multirow{2}{*}{\pass} & \multirow{2}{*}{\na} \\ \cline{3-3} & & Improving the performance by replacing public with external & & & & & & & \\ \hline
\multirow{2}{*}{75} & \multirow{2}{*}{\BP} & \tx{Token name} & \multirow{2}{*}{\pass} & \multirow{2}{*}{\na} & \multirow{2}{*}{\na} & \multirow{2}{*}{\na} & \multirow{2}{*}{\na} & \multirow{2}{*}{\na} & \multirow{2}{*}{\na} \\ \cline{3-3} & & Adding a token name variable for external apps & & & & & & & \\ \hline
\multirow{2}{*}{76} & \multirow{2}{*}{\BP} & \tx{Error information in revert condition} & \multirow{2}{*}{\na} & \multirow{2}{*}{\na} & \multirow{2}{*}{\na} & \multirow{2}{*}{\na} & \multirow{2}{*}{\pass} & \multirow{2}{*}{\na} & \multirow{2}{*}{\na} \\ \cline{3-3} & & Adding error description in require()/revert() to clarify the reason & & & & & & & \\ \hline
\multirow{2}{*}{77} & \multirow{2}{*}{\BP} & \tx{Complex Fallback} & \multirow{2}{*}{\na} & \multirow{2}{*}{\na} & \multirow{2}{*}{\na} & \multirow{2}{*}{\na} & \multirow{2}{*}{\pass} & \multirow{2}{*}{\na} & \multirow{2}{*}{\na} \\ \cline{3-3} & & Logging operations in the fallback() to avoid complex operations & & & & & & & \\ \hline
\multirow{2}{*}{78} & \multirow{2}{*}{\BP} & \tx{Function Order} & \multirow{2}{*}{\na} & \multirow{2}{*}{\na} & \multirow{2}{*}{\na} & \multirow{2}{*}{\na} & \multirow{2}{*}{\pass} & \multirow{2}{*}{\na} & \multirow{2}{*}{\na} \\ \cline{3-3} & & Following fallback, external, public, internal and private order & & & & & & & \\ \hline
\multirow{2}{*}{79} & \multirow{2}{*}{\BP} & \tx{Visibility Modifier Order} & \multirow{2}{*}{\na} & \multirow{2}{*}{\na} & \multirow{2}{*}{\na} & \multirow{2}{*}{\na} & \multirow{2}{*}{\na} & \multirow{2}{*}{\pass} & \multirow{2}{*}{\na} \\ \cline{3-3} & & Specifying visibility first and before modifiers in functions & & & & & & & \\ \hline
\multirow{2}{*}{80} & \multirow{2}{*}{\BP} & \tx{Non-initialized return value} & \multirow{2}{*}{\na} & \multirow{2}{*}{\pass} & \multirow{2}{*}{\na} & \multirow{2}{*}{\na} & \multirow{2}{*}{\pass} & \multirow{2}{*}{\na} & \multirow{2}{*}{\na} \\ \cline{3-3} & & Not specifying return for functions without output & & & & & 6& & \\ \hline
\multirow{2}{*}{81} & \multirow{2}{*}{\BP} & \tx{Token symbol} & \multirow{2}{*}{\pass} & \multirow{2}{*}{\na} & \multirow{2}{*}{\na} & \multirow{2}{*}{\na} & \multirow{2}{*}{\na} & \multirow{2}{*}{\na} & \multirow{2}{*}{\na} \\ \cline{3-3} & & Adding token symbol variable for usage of external apps & & & & & & & \\ \hline
\multirow{2}{*}{82} & \multirow{2}{*}{\BP} & \tx{Allowance spending is possible} & \multirow{2}{*}{\pass} & \multirow{2}{*}{\na} & \multirow{2}{*}{\na} & \multirow{2}{*}{\na} & \multirow{2}{*}{\na} & \multirow{2}{*}{\na} & \multirow{2}{*}{\na} \\ \cline{3-3} & & Ability of token transfer by transferFrom() to transfer tokens on behalf of another usercalc & & & & & & & \\ \hline

\hline
\rcl
\multicolumn{3}{|c|}{\ccl\tx{\begin{tabular}[c]{@{}c@{}}\prct success rate in performed audits by considering\\ 'False Positives' and 'Informational' checks as 'Passed' \\ (More details in section\ref{sec:tools}) \end{tabular}}}  & 100\% & 100\% & 100\% & 100\% & 100\% & 100\% & 97\% \\ \hline

\end{tabular}
\end{adjustbox}	
\caption{Continuation of Table\ref{tab:result2}.\label{tab:result3}}
\end{table*}

\section{Auditing Tools and \erc}\label{sec:tools}
Finally, we conducted an experiment on code auditing tools using the Solidity implementation of \sys to understand the current state of automated volunerabiliy testing. Our results illuminate the (in)completeness and error-rate of such tools on one specific use-case (related work studies, in greater width and less depth, a variety of use-cases~\cite{AuditTools}). We did not adapt older tools that support significantly lower versions of the Solidity compiler (\eg Oyente). We concentrated on Solidity as Vyper analysis is currently a paid services or penciled in for future support (\eg Slither). The provided version number is based on the GitHub repository; tools without a version are web-based and were used in 2020:

\begin{enumerate}
	\item EY Smart Contract \& Token Review by Ernst \& Young Global Limited~\cite{EYTool}.
	\item SmartCheck by SmartDec~\cite{SMARTCHECK}.
	\item Securify v2.0 by ChainSecurity~\cite{SECURIFYGIT,SECURIFY}.
	\item ContractGuard by GuardStrike~\cite{ContractGuard}.
	\item MythX by ConsenSys~\cite{MythX}.
	\item Slither Analyzer v0.6.12 by Crytic~\cite{SlitherDoc}.
	\item Odin by Sooho~\cite{Odin}.
\end{enumerate}

\subsection{Analysis of audit results}
A total of \num audits have been conducted by these auditing tools that are summarized in Tables~\ref{tab:result1}, \ref{tab:result2} and \ref{tab:result3}. Audits include best practices and security vulnerabilities. To compile the list of \num, we referenced the knowledge-base of each tool~\cite{SECURIFYGIT,SMARTCHECK,MythX,ContractGuard,SlitherDoc}, understood each threat, manually mapped the audit to the corresponding SWC registry~\cite{SWC}, and manually determined when different tools were testing for the same vulnerability or best practice (which was not always clear from the tools' own descriptions). Since each tool employs different methodology to analyze smart contracts (\eg comparing with violation patterns, applying a set of rules, using static analysis, \etc), there are false positives to manually check. Many false positives are not simply due to old/unmaintained rules but actually require tool improvement. We provide some examples in this section.

\textit{MythX} detects \textit{Re-entrancy attack} in the \textit{noReentrancy} modifier. In Solidity, modifiers are not like functions. They are used to add features or apply some restriction on functions~\cite{SolidityModifer}. Using modifiers is a known technique to implement Mutex and mitigate re-entrancy attack~\cite{ReentrancyGuard}. This is a false positive and note that other tools have not identified the attack in modifiers.

\textit{ContractGuard} flags \textit{Re-entrancy attack} in \texttt{transfer()} function while countermeasures (based on both CEI and Mutex~\ref{subsec:ent}) are implemented.

\textit{Slither} detects two \textit{low level call} vulnerabilities~\cite{SlitherSetup}. This is due to use of \texttt{call.value()} that is recommend way of transferring ETH after \textit{Istanbul} hard-fork (EIP-1884).	Therefore, adapting analyzers to new standards can improve accuracy of the security checks.

\textit{SmartCheck} recommends not using \texttt{SafeMath} and check explicitly where overflows might be occurred. We consider this failed audit as false possible whereas utilizing \texttt{SafeMath} is a known technique to mitigate over/under flows. It also flags \textit{using a private modifier} as a vulnerability by mentioning, ``miners have access to all contracts' data and developers must account for the lack of privacy in Ethereum''. However private visibility in Solidity concerns object-oriented inheritance not confidentiality. For actual confidentiality, the best practice is to encrypt private data or store them off-chain. The tool also warns against \texttt{approve()} in \erc due to \textit{front-running attacks}. Despite EIP-1884, it still recommends using of \texttt{transfer()} method with stipend of 2300 gas. There are other false positives such as SWC-105 and SWC-112 that are passed by other tools.

\textit{Securify} detects the \textit{Re-entrancy} attack due to unrestricted writes in the \texttt{noReentrancy} modifier~\cite{SECURIFY}. Modifiers are the recommended approach and are not accessible by users. It also flags \textit{Delegatecall to Untrusted Callee (SWC-112)} while there is no usage of \texttt{delegatecall()} in the code. It might be due to use of \texttt{SafeMath} library which is an embedded library. In Solidity, embedded libraries are called by JUMP commands instead of \texttt{delegatecall()}. Therefore, excluding embedded libraries from this check might improve accuracy of the tool. Similar to \textit{SmartCheck}, it still recommends to use the \texttt{transfer()} method instead of \texttt{call.value()}.

\textit{EY token review} considers \texttt{decreaseAllowance} and \texttt{increaseAllowance} as standard \erc functions and if not implemented, recognizes the code as vulnerable to a \textit{front-running}. These two functions are not defined in the \erc standard~\cite{ERC20Std} and considered only by this tool as mandatory functions. There are other methods to prevent the attack while adhering \erc specifications (see Rahimian \etal for a full paper on this attack and the basis of the mitigation in \sys~\cite{ERC20MWA}). The tool also falsely detects the \textit{Overflow}, mitigated through \texttt{SafeMath}. Another identified issue is \textit{Funds can be held only by user-controlled wallets}. The tool warns against any token transfer to Ethereum addresses that belong to smart contracts. However, interacting with \erc token by other smart contracts was one of the main motivations of the standard. It also checks for maximum 50000 gas in \texttt{approve()} and 60000 in \texttt{transfer()} method. We could not find corresponding SWC registry or standard recommendation on these limitations and therefore consider them as informational.

\textit{Odin} raises \textit{Outdated compiler version} issue due to locking solidity version to 0.5.11. We have used this version due to its compatibility with other auditing tools.


\begin{table}[t!]
\centering
\scalebox{0.85}{
\begin{tabular}{|l|c|c|c|c|c|c|c|c|}
	\rcl
	\multicolumn{1}{|c|}{\ccl} & \multicolumn{7}{c|}{\ccl\tx{Auditing Tool}} & \ccl \\ \cline{2-8}
	\rcl 
	\multicolumn{1}{|c|}{\multirow{-2}{*}{\ccl\tx{\begin{tabular}[c]{@{}c@{}}ERC-20\\ Token\end{tabular}}}} & \begin{tabular}[c]{@{}c@{}}EY Token\\ Review\end{tabular} & \begin{tabular}[c]{@{}c@{}}Smart\\ Check\end{tabular} & Securify & \begin{tabular}[c]{@{}c@{}}MythX\\ (Mythril)\end{tabular} & \begin{tabular}[c]{@{}c@{}}Contract\\ Guard\end{tabular} & Slither & Odin & \multirow{-2}{*}{\ccl\tx{\begin{tabular}[c]{@{}c@{}}Total\\ issues\end{tabular}}} \\ \hline
	\tx{\sys} & 9 & 11 & 4 & 2 & 10 & 2 & 2 & \tx{40} \\ \hline
	\tx{TUSD} & 20 & 11 & 2 & 1 & 14 & 16 & 6 & \tx{70} \\ \hline
	\tx{PAX} & 16 & 9 & 6 & 4 & 16 & 13 & 9 & \tx{73} \\ \hline
	\tx{USDC} & 17 & 9 & 6 & 5 & 18 & 15 & 10 & \tx{80} \\ \hline
	\tx{INO} & 11 & 10 & 14 & 8 & 14 & 24 & 12 & \tx{93} \\ \hline
	\tx{HEDG} & 10 & 28 & 11 & 1 & 29 & 24 & 16 & \tx{119} \\ \hline
	\tx{BNB} & 13 & 21 & 12 & 13 & 41 & 39 & 3 & \tx{142} \\ \hline
	\tx{MKR} & 11 & 27 & 38 & 9 & 16 & 34 & 18 & \tx{153} \\ \hline
	\tx{LINK} & 12 & 27 & 38 & 9 & 16 & 34 & 18 & \tx{181} \\ \hline
	\tx{USDT} & 12 & 29 & 8 & 17 & 46 & 55 & 30 & \tx{197} \\ \hline
	\tx{LEO} & 32 & 25 & 8 & 23 & 70 & 75 & 19 & \tx{252} \\ \hline
\end{tabular}}
\caption{Security flaws detected by seven auditing tools in \sys (the proposal) compared to top 10 \erc tokens by market capitalization in May 2020. \sys has the lowest reported security issues (occurrences). \label{tab:summary}}
\end{table}

\subsection{Comparing audits}
After manually overriding the false positives, the average percentage of passed checks for \sys reaches to \prct. To pass the one missing check and reach a 100\% success rate across all tools, we prepared the same code in Solidity version 0.8.4, however it cannot be audited anymore with most of the tools. 

We repeated the same auditing process on the top ten tokens based on their market cap~\cite{EtherScan}. The result of all these evaluation have been summarized in Table~\ref{tab:summary} by considering false positives as failed audits. This provides the same evaluation conditions across all tokens. Since each tool uses different analysis methods, number of occurrences are considered for comparisons. For example, MythX detects two \textit{re-entrancy} in \sys; therefore, two occurrences are counted instead of one. 

As it can be seen in Table~\ref{tab:summary}, \sys has the least number of security flaws (occurrences) compared to other tokens. We stress that detected security issues for \sys are all false positives. We are also up-front that this metric is not a perfect indication of security.  The other tokens may also have many/all false positives (such an analysis would be interesting future work), and not all true positives can be exploited~\cite{VulExp}. Mainly, we want to show this measurement as being consistent with our claims around the security of \sys. Had \sys, for example, had the highest number of occurrences, it would be a major red flag.


\section{Conclusion}

98\% of tokens on Ethereum today implement \erc. While attention has been paid to the security of Ethereum DApps, threats to tokens can be specific to \erc functionality. In this paper, we provide a detailed study of \erc security, collecting and deduplicating applicable vulnerabilities and best practices, examining the ability of seven audit tools. Most importantly, we provide a concrete implementation of \erc called \sys\footnote{Compatible Solidity version of \sys (v. 0.5.11) deployed on Mainnet at \url{https://bit.ly/35FMbAf} and the latest Solidity (v. 0.8.4) on Rinkeby \url{https://bit.ly/3tI139S}. Vyper code at \url{https://bit.ly/3dXaaPc}.}. It is designed to be secure against known vulnerabilities, and can serve as a second reference implementation to provide software diversity. We test it at Solidity version 0.5.11 (due to the limitation of the audit tools) and also provide it at version 0.8.4. Vyper implementation is also provided at version 0.2.8 to make \erc contracts more secure and easier to audit. \sys can be used as template to deploy new \erc tokens (\eg ICOs, DApps, etc), migrate current vulnerable deployments, and to benchmark the precision of Ethereum audit tools.




{\footnotesize\bibliography{bib/references}}


\end{document}